\documentclass[aps,pra,reprint,twocolumn,superscriptaddress,floatfix,nofootinbib,longbibliography]{revtex4-2}
\usepackage{graphicx,amsmath,amsfonts,amssymb,xr}
\usepackage{epsfig, color, dsfont, upgreek, physics}
\usepackage{mathrsfs}
\usepackage{mathtools}
\usepackage{bbold}
\usepackage{comment}
\usepackage{braket}
\usepackage{lipsum}
\usepackage{soul}
\usepackage{bm}

\usepackage[bookmarks=true,colorlinks,linkcolor=OrangeRed,urlcolor=NavyBlue,citecolor=RoyalBlue]{hyperref}
\usepackage[capitalise]{cleveref}
\Crefname{figure}{Fig.}{Figs.}
\Crefname{section}{Sec.}{Secs.}
\usepackage[dvipsnames]{xcolor}
\usepackage[inline]{enumitem}
\setlist[enumerate]{label=(\roman*)}
\usepackage{cmds}
\graphicspath{{figures/}}
\usepackage{stackengine, scalerel}

\usepackage{orcidlink}

\newcommand{\orcidjad}{\orcidlink{0000-0002-0659-7990}}
\newcommand{\orcidsimone}{\orcidlink{0009-0006-3594-8100}}
\newcommand{\orcidguoxian}{\orcidlink{0000-0001-7936-762X}}
\newcommand{\orcidbing}{\orcidlink{0000-0002-8379-9289}}
\newcommand{\LMU}{\affiliation{Department of Physics and Arnold Sommerfeld Center for Theoretical Physics (ASC), Ludwig Maximilian University of Munich, 80333 Munich, Germany}}
\newcommand{\MPQ}{\affiliation{Max Planck Institute of Quantum Optics, 85748 Garching, Germany}}
\newcommand{\MCQST}{\affiliation{Munich Center for Quantum Science and Technology (MCQST), 80799 Munich, Germany}}
\newcommand{\UNITO}{\affiliation{Dipartimento di Fisica, Università di Torino, I-10125 Torino, Italy}}
\newcommand{\KYUNGHEE}{\affiliation{Department of Physics, College of Science, Kyung Hee University, Seoul 02447, Republic of Korea}}
\newcommand{\MIT}{\affiliation{Research Laboratory of Electronics, MIT-Harvard Center for Ultracold Atoms, Department of Physics, Massachusetts Institute of Technology, MA 02139, USA}}
\newcommand{\SUST}{\affiliation{Department of Physics, Southern University of Science and Technology, Shenzhen 518055, China}}

\newcommand{\roundbr}[1]{\left( #1 \right)}
\newcommand{\rb}{\textbf{r}}
\newcommand{\jb}{\textbf{j}}
\newcommand{\m}{m_{\text{eff}}}
\newcommand{\Hspace}{\mathcal{H}}
\newcommand{\Proj}{\mathcal{P}}

\begin{document}

\title{Scalable cold-atom quantum simulator of a $3+1$D U$(1)$ lattice gauge theory with dynamical matter}
\author{Simone Orlando$^{\orcidsimone}$} \UNITO \MPQ \MCQST
\author{Guo-Xian Su$^{\orcidguoxian}$} \MIT 
\author{Bing Yang$^{\orcidbing}$} \SUST
\author{Jad C.~Halimeh$^{\orcidjad}$}\email{jad.halimeh@lmu.de}\LMU \MPQ \MCQST \KYUNGHEE
\date{\today}
\begin{abstract}
The stated overarching goal of the highly active field of quantum simulation of high-energy physics (HEP) is to achieve the capability to study \textit{ab-initio} real-time microscopic dynamics of $3+1$D quantum chromodynamics (QCD). However, existing experimental realizations and theoretical proposals for future ones have remained restricted to one or two spatial dimensions. Here, we take a big step towards this goal by proposing a concrete experimentally feasible scalable cold-atom quantum simulator of a U$(1)$ quantum link model of quantum electrodynamics (QED) in three spatial dimensions, employing \textit{linear gauge protection} to stabilize gauge invariance. Using tree tensor network simulations, we benchmark the performance of this quantum simulator through near- and far-from-equilibrium observables, showing excellent agreement with the ideal gauge theory. Additionally, we introduce a method for \textit{analog quantum error mitigation} that accounts for unwanted first-order tunneling processes, vastly improving agreement between quantum-simulator and ideal-gauge-theory results. Our findings pave the way towards realistic quantum simulators of $3+1$D lattice gauge theories that can probe regimes well beyond  classical simulability.
\end{abstract}

\maketitle
\tableofcontents
\section{Introduction}
Gauge theories form the foundation of modern physics, encoding the fundamental laws of nature through the principle of local symmetry, and constitute the Standard Model of particle physics \cite{Weinberg1995QuantumTheoryFields,Gattringer2009QuantumChromodynamicsLattice,Zee2003QuantumFieldTheory}. Their lattice variants, lattice gauge theories (LGTs) \cite{Rothe2012LatticeGaugeTheories}, were conceived to study quark confinement \cite{Wilson1974ConfinementQuarks,Wilson1977QuarksStringsLattice} but, in recent years, they have been used in fields beyond high-energy physics (HEP). For example, they are powerful tools for describing emergent exotic phases in condensed matter systems such as quantum spin liquids and frustrated magnets \cite{Wegner1971DualityInGeneralizedIsingModels,Kogut1979AnIntroductionToLatticeGaugeTheory,wen2004quantum,Sedgewick2002FractionalizedPhase,Sachdev2016TheNovelMetallicStates} and have been used in theoretical investigations of high-temperature superconductivity \cite{Lee2008FromHighTemperatureSuperconductivity,Senthil20002GaugeTheory}. In quantum many-body physics, they have been a major venue for nonergodic dynamical phenomena such as quantum many-body scarring \cite{Turner2018WeakErgodicityBreaking,Moudgalya2018ExactExcitedStates,Surace2020LatticeGaugeTheories,Iadecola2020QuantumManyBodyScar,Zhao2020QuantumManyBodyScars,Aramthottil2022ScarStates,Biswas2022ScarsFromProtectedZeroModes,Jepsen2022Long-LivedPhantomHelix,Serbyn2021QuantumManyBodyScars,Moudgalya2022QuantumManyBodyScarsHilbertSpaceFragmentation,Chandran2023QuantumManyBodyScars,Bluvstein2022QuantumProcessor,Desaules2023WeakErgodicityBreaking,Desaules2023ProminentQuantumManyBodyScars,Zhang2023ManyBodyHilbertSpaceScarring,Dong2023DisorderTunableEntanglement,Osborne2024QuantumManyBodyScarring,Budde2024QuantumManyBodyScars,Hartse2025StabilizerScars} and disorder-free localization \cite{Smith2017DisorderFreeLocalization,Brenes2018ManyBodyLocalization,Smith2017AbsenceOfErgodicity,Karpov2021DisorderFreeLocalization, Sous2021PhononInducedDisorder,Chakraborty2022DisorderFreeLocalization,Halimeh2022EnhancingDisorderFreeLocalization,Homeier2023RealisticScheme,Osborne2023DisorderFreeLocalization$2+1$D,Cataldi2025DisorderFreeLocalizationFragmentation-1}.

Their utility across various disciplines has made them a very attractive choice for quantum simulation \cite{Byrnes2006SimulatingLatticeGauge, Dalmonte2016LatticeGaugeTheory, Zohar2015QuantumSimulationsLattice, Aidelsburger2021ColdAtomsMeet, Zohar2021QuantumSimulationLattice, Klco2022StandardModelPhysics, Bauer2023QuantumSimulationHighEnergy, Bauer2023QuantumSimulationFundamental,
DiMeglio2024QuantumComputingHighEnergy, Cheng2024EmergentGaugeTheory, Halimeh2022StabilizingGaugeTheories, Halimeh2023ColdatomQuantumSimulators, Cohen2021QuantumAlgorithmsTransport, Lee2025QuantumComputingEnergy, Turro2024ClassicalQuantumComputing, Bauer2025EfficientUseQuantum}. Such quantum simulations facilitate addressing open questions at the heart of HEP, condensed matter, and quantum many-body physics, such as the fate of an isolated interacting quantum system quenched far from equilibrium under gauge-symmetry constraints. Not only is this pertinent to the Eigenstate Thermalization Hypothesis \cite{Deutsch1991QuantumStatisticalMechanics,Srednicki1994ChaosAndQuantumThermalization,Rigol2008Thermalization,Eisert2015QuantumManyBodySystems,DAlessio2016FromQuantumChaos,Kaufman2016QuantumThermalization,Deutsch2018ETH}, but it directly touches on how the early universe evolved \cite{mukhanov2005physical,weinberg2008cosmology}. This has spurred a major effort in the field that has led to impressive quantum-simulation experiments of LGTs on various quantum-hardware platforms \cite{Martinez2016RealtimeDynamicsLattice, Klco2018QuantumclassicalComputationSchwinger,Gorg2019RealizationDensitydependentPeierls, Schweizer2019FloquetApproachZ2, Mil2020ScalableRealizationLocal, Yang2020ObservationGaugeInvariance, Wang2022ObservationEmergent$mathbbZ_2$, Su2023ObservationManybodyScarring, Zhou2022ThermalizationDynamicsGauge, Wang2023InterrelatedThermalizationQuantum, Zhang2025ObservationMicroscopicConfinement, Zhu2024ProbingFalseVacuum, Ciavarella2021TrailheadQuantumSimulation, Ciavarella2022PreparationSU3Lattice, Ciavarella2023QuantumSimulationLattice-1, Ciavarella2024QuantumSimulationSU3, 
Gustafson2024PrimitiveQuantumGates, Gustafson2024PrimitiveQuantumGates-1, Lamm2024BlockEncodingsDiscrete, Farrell2023PreparationsQuantumSimulations-1, Farrell2023PreparationsQuantumSimulations, 
Farrell2024ScalableCircuitsPreparing,
Farrell2024QuantumSimulationsHadron, Li2024SequencyHierarchyTruncation, Zemlevskiy2025ScalableQuantumSimulations, Lewis2019QubitModelU1, Atas2021SU2HadronsQuantum, ARahman2022SelfmitigatingTrotterCircuits, Atas2023SimulatingOnedimensionalQuantum, Mendicelli2023RealTimeEvolution, Kavaki2024SquarePlaquettesTriamond, Than2024PhaseDiagramQuantum, Angelides2025FirstorderPhaseTransition, Gyawali2025ObservationDisorderfreeLocalization, Cochran2025VisualizingDynamicsCharges, Gonzalez-Cuadra2025ObservationStringBreaking, Crippa2024AnalysisConfinementString, De2024ObservationStringbreakingDynamics, Liu2024StringBreakingMechanism, Alexandrou2025RealizingStringBreaking, 
Mildenberger2025Confinement$$mathbbZ_2$$Lattice, Schuhmacher2025ObservationHadronScattering, Davoudi2025QuantumComputationHadron, Cobos2025RealTimeDynamics2+1D, Saner2025RealTimeObservationAharonovBohm, Xiang2025RealtimeScatteringFreezeout, Wang2025ObservationInelasticMeson,li2025frameworkquantumsimulationsenergyloss,froland2025simulatingfullygaugefixedsu2,Hudomal2025ErgodicityBreakingMeetsCriticality}. 

Despite this great progress, all these experimental works have been restricted to LGTs in one or two spatial dimensions, but three spatial dimensions\footnote{We denote a system or phenomenon in $d$ spatial dimensions as $d+1$D or $d$d.} have so far been elusive. This is a major challenge where even theoretical proposals have been lacking, but is of utmost importance to model the physical world. Indeed, a stated goal of the field of quantum simulation for HEP is to eventually quantum-simulate the microscopic dynamics of $3+1$D QCD, elevating quantum simulators of HEP to venues complementary to dedicated particle colliders \cite{Bauer2023QuantumSimulationFundamental,DiMeglio2024QuantumComputingHighEnergy}. This is also crucial in condensed matter and quantum many-body dynamics. It is known that quantum criticality in and out of equilibrium directly depends on spatial dimensions. Powerful numerical techniques such as tensor networks \cite{Montangero2018IntroductionTensorNetwork,Schollwock2011DensitymatrixRenormalizationGroup,Paeckel2019TimeevolutionMethodsMatrixproduct,Orus2019TensorNetworksComplex} fundamentally struggle in higher spatial dimensions due to the rapid buildup of quantum entanglement. Particle colliders naturally accommodate $3+1$D gauge theories, but extracting snapshots of microscopic dynamics at arbitrary evolution times is quite challenging \cite{Ellis2003QCDColliderPhysics}. Monte Carlo methods can handle higher spatial dimensions, but suffer from the sign problem out of equilibrium \cite{deforcrand2010simulatingqcdfinitedensity,Troyer2005ComputationalComplexityFundamental}. This motivates the pursuit of a robust quantum simulator of a $3+1$D LGT that can constitute a first step towards probing $3+1$D QCD on quantum hardware.

In this work, we propose an experimentally feasible scalable cold-atom quantum simulator of a $3+1$D quantum link model (QLM) \cite{Chandrasekharan1997QuantumLinkModels,wiese_ultracold_2013,kasper_implementing_2017} formulation of QED.
The ingredients of this model can be readily implemented using existing techniques integrated into the system, particularly for state initialization, quantum-state manipulation, and detection.
The quantum simulator stabilizes gauge invariance through linear gauge protection, a scheme based on quantum Zeno dynamics \cite{Halimeh2022StabilizingGaugeTheories,Halimeh2021GaugeSymmetryProtection}. We benchmark the efficacy of our proposed quantum simulator through tree tensor network (TTN) \cite{tagliacozzo_simulation_2009,murg_simulating_2010} simulations, showing excellent agreement with the ideal-LGT results in and out of equilibrium.

\begin{figure*}
    \centering
    \includegraphics[width=\linewidth]{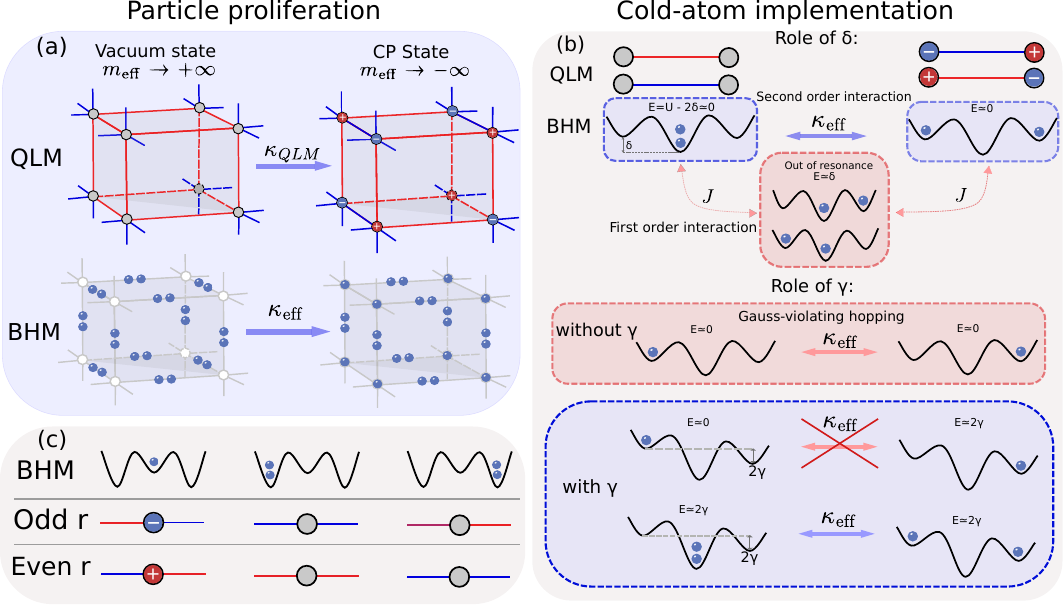}
    \caption{Schematic for our proposed scalable cold-atom analog quantum simulator. \textbf{(a)} Example of a transition between a zero-particle (bare) vacuum state to a charge-proliferated one in the QLM and BHM representations. We study this transition by varying the mass with a ramping protocol as reported in Fig.~\ref{fig:Ramp dynamics}. \textbf{(b)} We show how to simulate the desired dynamics in cold atoms trapped in an optical superlattice. The $\delta$ term in Hamiltonian \eqref{eq:BHM_hamiltonian} restricts the Hilbert space in the desired subspace required to faithfully capture the QLM dynamics. The $\gamma$ term creates a tilt in potential and is used to prevent dynamics that would violate Gauss law. \textbf{(c)} Mapping between the BHM and the QLM. Using the staggered-fermion representation, a single atom on an odd (even) site represents a ``$-$'' (``$+$'') particle (antiparticle), which can be viewed as an electron (positron). The red (blue) links represent eigenstates of $\hat S^z$ with eigenvalue $\pm\frac{1}{2}$, which correspond to double (zero) occupied link sites in the BHM, respectively (see text).}
    \label{fig:Figure1}
\end{figure*}

\section{Target $3+1$D lattice gauge theory}

We consider the U$(1)$ LGT \cite{kogut_hamiltonian_1975}
\begin{align}\nonumber
        \hat H_{\text{U(1)}} = &- \kappa \sum_{\rb, \mu} \roundbr{\hat \phi_\rb^\dagger \hat U_{\rb, \rb +\mu} \hat \phi_{\rb+\mu} + \text{H.c.}} \\\nonumber
        &+m\sum_\rb (-1)^\rb \hat \phi_\rb^\dagger\hat \phi_\rb + g^2\sum_{\rb, \mu} \hat E_{\rb, \rb+\mu}^2\\ 
    \label{eq:Gauge_Hamiltonian}
    &+ \frac{1}{g^2}\sum_\square \roundbr{\hat U_\square + \hat U_\square^\dagger},
\end{align}
where $\textbf{r} = (r_x, r_y, r_z) \in \mathbb{N}^3$ represents the coordinate of matter sites in a cubic $3$d lattice space $\Lambda$ with unity lattice spacing, $\mu$ represents a unit vector along one direction $\{x,y,z\}$, $\kappa$ is the gauge-invariant hopping, where the minimal coupling to the U$(1)$ gauge field is mediated by the link operator $\hat{U}_{\mathbf{r}, \mathbf{r}+\mu}$ residing on the bond between sites $\mathbf{r}$ and $\mathbf{r}+\mu$, which also hosts the electric-field operator $\hat{E}_{\mathbf{r}, \mathbf{r}+\mu}$, $m$ is the rest mass of the matter field, with the staggering according to the parity of site $\mathbf{r}$ is to distinguish between (anti)particle on even (odd) sites \cite{Susskind1977LatticeFermions}, and $g$ is the gauge coupling. Here, we consider hardcore-bosonic matter fields satisfying the canonical relations $\big\{\hat \phi_\textbf{r}, \hat \phi_{\textbf{r}'}^\dagger\big\} = \delta_{\textbf{r},\textbf{r}'}$ and $\big[\hat \phi_\textbf{r}, \hat \phi_{\textbf{r}'}^\dagger\big] = \delta_{\textbf{r},\textbf{r}'}\big(1-2\hat\phi_\textbf{r}^\dagger\hat\phi_\textbf{r}\big)$. The plaquette operator $\hat{U}_{\square} $, which is a product of all the link operators $\hat U_{\mathbf{r}, \mathbf{r}+\mu}$ over a minimal plaquette. This encodes the magnetic field energy through the Wilson formulation $\hat{U}_{\square} + \hat{U}^\dagger_{\square}$. The gauge- and electric-field operators satisfy the canonical commutation relations 
\begin{subequations}\label{eq:CanCom}
\begin{align}
&\big[\hat{U}_{\mathbf{r},\mathbf{r}+\mu}, \hat{U}_{\mathbf{r}',\mathbf{r}'+\mu'}^\dagger\big]=0,\\
&\big[\hat{E}_{\mathbf{r},\mathbf{r}+\mu}, \hat{U}_{\mathbf{r}',\mathbf{r}'+\mu'}\big] = \delta_{\mathbf{r},\mathbf{r}'}\delta_{\mu,\mu'} \hat{U}_{\mathbf{r},\mathbf{r}+\mu}.
\end{align}    
\end{subequations}
Hamiltonian \eqref{eq:Gauge_Hamiltonian} is invariant with respect to the U$(1)$-gauge-symmetry generator
 \begin{align}
     \hat G_\rb = \sum_\mu \left(\hat E_{\rb, \rb + \mu} - \hat E_{\rb - \mu, \rb }\right) + \frac{1- (-1)^\rb}{2} - \hat\phi^\dagger_\rb \hat \phi_{\rb},
 \end{align}
which stipulates an intrinsic relationship between matter on a given site and the allowed electric-field configurations on its six neighboring links. Gauge-invariant states $\ket{\Psi}$ are simultaneous eigenstates of all generators, satisfying $\hat{G}_\rb\ket{\Psi}=g_\rb\ket{\Psi},\,\forall\rb$. The eigenvalues $g_\rb$ are so-called background charges, and a set of them over the entire lattice defines a gauge sector. The \textit{physical} sector, also known as the sector of Gauss's law, is that where $g_\rb=0,\,\forall\rb$.

The gauge and electric fields in Hamiltonian \eqref{eq:Gauge_Hamiltonian} are infinite-dimensional, rendering its implementation intractable on existing quantum hardware and numerical frameworks. To facilitate both numerical and quantum simulations, we use a QLM representation \cite{Chandrasekharan1997QuantumLinkModels,wiese_ultracold_2013,kasper_implementing_2017} of the U$(1)$ LGT, where we substitute the electric-flux operator $\hat E_{\rb, \rb + \mu}$ and the gauge-field operator $\hat U_{\rb, \rb + \mu}$  with spin-$S$ operators:
\begin{subequations}
    \begin{align}
        \hat U_{\rb, \rb + \mu} &\to \frac{\hat S^+_{\rb, \rb + \mu}}{\sqrt{S(S+1)}}, \\
        \hat E_{\rb, \rb + \mu} &\to \hat S^z_{\rb, \rb + \mu}.
    \end{align}
\end{subequations}
In the Kogut--Susskind limit of $S\to \infty$, this representation preserves the canonical commutation relations~\eqref{eq:CanCom} \cite{buyens_finite-representation_2017, carmen_banuls_review_2020, zache_toward_2022, halimeh_achieving_2022}.

We further perform the particle-hole transformation \cite{Hauke2013QuantumSimulationLattice,osborne_large-scale_2025, yang_analog_2016} 

\begin{subequations}
    \begin{align}
        \hat \phi_\rb &\to \frac{1+(-1)^\rb}{2}\hat \phi_\rb+\frac{1-(-1)^\rb}{2}\hat \phi_\rb^\dagger,\\
        \hat \phi_\rb^\dagger \hat \phi_\rb&\to\frac{1-(-1)^\rb}{2}+(-1)^\rb\hat \phi_\rb^\dagger \hat \phi_\rb,\\
        \hat S^+_{\rb,\rb+\mu}&\to\frac{1-(-1)^\rb}{2}\hat S^+_{\rb,\rb+\mu}+\frac{1+(-1)^\rb}{2}\hat S^-_{\rb,\rb+\mu},\\
        \hat S^z_{\rb,\rb+\mu} &\to (-1)^{\rb+\mu} \hat S^z_{\rb,\rb+\mu}.
    \end{align}
\end{subequations}
We consider a high-coupling regime $g^2 \to \infty$, which is equivalent to taking $S=\frac{1}{2}$. In this regime, the plaquette term is also negligible. Our target Hamiltonian is now
\begin{align}\nonumber
    \hat H_{\text{target}} = &-\kappa_{\text{QLM}}\sum_{\textbf{r}, \mu}  \left( \hat \phi^\dagger_\rb \hat S^-_{\textbf{r},\textbf{r}+\mu} \hat \phi^\dagger_{\textbf{r} +\mu} + \rm{H.c.}\right)\\    \label{eq:QLMhamiltonian}
    &+ m\sum_\rb \hat \phi^\dagger_\textbf{r} \hat \phi_\textbf{r}.
\end{align}
Note that since we are in a spin-$\frac{1}{2}$ representation, the gauge-coupling term $\propto \left(\hat S^z_{\textbf{r},\textbf{r}+\mu}\right)^2$ becomes an inconsequential energetic constant. The generator of the U$(1)$ gauge symmetry associated with Hamiltonian \eqref{eq:QLMhamiltonian} is
\begin{align}
\hat{G}_\textbf{r} = (-1)^{r_x + r_y + r_z} \left[ \hat{ \phi}_\rb^\dagger \hat{\phi}_\rb + \sum_{\mu} \left( \hat{s}_{\rb, \rb +\mu}^z + \hat{s}_{\rb-\mu,\rb}^z \right) \right],
\end{align}
where gauge invariance is encoded in the condition $[\hat{H}_{\text{target}}, \hat{G}_\rb] = 0,\,\forall\rb$. In the following, we want to work in the physical sector, i.e., $\hat{G}_\rb |\Psi\rangle = 0,\,\forall\rb$. 

\section{Mapping to $3+1$D Bose--Hubbard model}
We now map the target LGT~\eqref{eq:QLMhamiltonian} to a $3$d Bose--Hubbard model (BHM). The latter is the standard effective model for interacting bosons in optical lattices. In cold-atom experiments, bosonic atoms in a sufficiently deep optical lattice naturally realize its key ingredients: nearest-neighbor tunneling, on-site interaction, and chemical potential \cite{Bloch2008ManyBodyPhysics,Gross2017QuantumSimulations,Chanda2025_review}. We consider a new $3$d cubic lattice $\tilde\Lambda$ with sites $\jb$, and we identify even sites $\jb = 2\rb$) as \textit{matter site}. For each of these sites the adjacent (odd) sites $\jb = 2\rb +\tilde\mu$---here, $\tilde\mu\parallel\mu$ is a unit directional vector on $\tilde\Lambda$---are considered \textit{link sites}, and all the others sites of the lattice are \textit{forbidden}, in that they should take no part in the LGT physics. The key to this mapping lies in appropriately restricting the local Hilbert spaces of BHM \cite{Yang2020ObservationGaugeInvariance}. At each matter site, we restrict the local Hilbert space to two states, $\Hspace_\jb = \{\ket 0_\jb, \ket 1_\jb\}$, representing the absence and presence of matter on that site. Similarly, at each link site, we restrict $\Hspace_\jb = \{\ket 0_\jb, \ket 2_\jb\}$, corresponding to the two possible orientations of the electric field on that link. We then define the projector to $\Hspace_\jb$ as $\hat\Proj_\jb$. In this encoding it is easy to shown that the bosonic creation (annihilation) operators $\hat b_\jb^\dagger(\hat b_\jb)$ mimic the commutation relations of the spin operators $S^{\pm}_{\rb, \rb + \mu}$:
\begin{subequations}\label{eq:mapping}
    \begin{align}
        \hat\phi^{(\dagger)}_\rb &= \hat\Proj_{2\rb} \hat b_{2\rb}^{(\dagger)} \hat\Proj_{2\rb},\\
        \hat\phi^{\dagger}_\rb \hat\phi_\rb &= \hat\Proj_{2\rb} \hat b_{2\rb}^{\dagger} b_{2\rb}\hat\Proj_{2\rb},\\
        \hat S^{-(+)}_{\rb + \mu} &= \frac{1}{\sqrt{2}} \hat\Proj_{2\rb+\tilde\mu}\left( \hat b_{2\rb+\tilde\mu}^{(\dagger)}\right)^2 \hat\Proj_{2\rb+\tilde\mu},\\
        \hat S^{z}_{\rb + \mu} &= \frac{1}{2} \hat\Proj_{2\rb+\tilde\mu} \left( \hat b_{2\rb+\tilde\mu}^{\dagger} \hat b_{2\rb+\tilde\mu}-1\right)\hat\Proj_{2\rb+\tilde\mu}.
    \end{align}
\end{subequations}
An empty link site corresponds to a local electric field eigenvalue $S_{r,e_\nu}^z = -\frac{1}{2}$. Conversely, a doublon in the bosonic model represents a local electric field eigenvalue $S_{r,e_\nu}^z=\frac{1}{2}$, see Fig.~\ref{fig:Figure1}.
Substituting Eqs.~\eqref{eq:mapping} in Hamiltonian \eqref{eq:QLMhamiltonian}, we obtain
\begin{align}\nonumber
    \hat H_{\text{eff}} = \sum_{\jb = 2\rb, \mu}\hat\Proj_\jb \hat\Proj_{\jb+\tilde\mu}\bigg[&\frac{\kappa_{\text{eff}}}{\sqrt{2}} \roundbr{\hat b_\jb^\dagger \hat b_{\jb +\tilde\mu}^2 \hat b_{\jb + 2 \tilde\mu}^\dagger+ \text{H.c.}}\\
    \label{eq:Effective hamiltonian}
    &+m \hat b_\jb^\dagger \hat b_\jb\bigg] \hat\Proj_\jb \hat\Proj_{\jb+\tilde\mu}.
\end{align}
To effectively implement (\ref{eq:Effective hamiltonian}) we use: 
\begin{align}\nonumber
        \hat H_{\rm{BHM}} = \sum_{\jb, \tilde\mu}\bigg[& -J\big(\hat b_\jb^\dagger \hat b_{\jb+\tilde\mu} + \text{H.c.}\big) +\frac{U_\jb}{2}\hat n_\jb (\hat n_\jb -1) \\\label{eq:BHM_hamiltonian}
        &+ \left(\boldsymbol{\gamma}\cdot \jb - \eta_\jb- \delta_\jb \right)\hat n_\jb\bigg].
    \end{align}
Where $J$ is the tunneling strength between nearest sites, $U_{\jb}$ is the on-site interaction that equals $U$ on link-sites and $2U$ on matter sites.
In order to impose the local Hilbert space constraints, we use a staggering term $\delta_j$ that equals zero on matter sites and $\delta$ on link sites. If we tune this staggering term such that $2\delta \simeq U$ then the energies $\hat H_{\text{BHM}}\ket{020} \simeq \hat H_{\text{BHM}}\ket{101}$ so that we have a second-order resonance between these states, see Fig.~\ref{fig:Figure1}b. 
Within the BHM formalism, The generator of Gauss's law takes the form
\begin{equation}
    \hat G_\jb =(-1)^{\jb} \bigg[\hat n_\jb + \frac{1}{2}\sum_{\tilde\mu} \hat n_{\jb + \tilde\mu}\left(\hat n_{\jb + \tilde\mu} -1\right) -3\bigg],
\end{equation}
where $\jb$ is here understood to be a matter site on $\tilde\Lambda$.
To constrain the system in the physical sector $\hat G_\jb \ket{\Psi} = 0,\, \forall \jb$, we introduce a linear gauge protection term \cite{Halimeh2021GaugeSymmetryProtection,Halimeh2020RobustnessOfGaugeInvariantDynamics}, which consists of a tilting potential in each direction $\boldsymbol{\gamma} = (\gamma_x, \gamma_y, \gamma_z)$, where $\gamma_i \neq \gamma_j,\, \forall i\neq j$; see Fig.~\ref{fig:Figure1}. We add a term $\eta_\jb = \eta>> \delta$ only on forbidden sites, and it is zero otherwise to prevent particles tunneling there. 
By means of almost degenerate second-order perturbation theory \cite{cohentannoudji_atomphoton_1998}, these terms restrict the effective dynamics within the physical sector and implement the desired effective Hamiltonian \eqref{eq:Effective hamiltonian} with $\kappa_{\text{eff}} \simeq 4 \sqrt{2} J^2/U$ and an effective mass $m_{\text{eff}}$ \cite{osborne_large-scale_2025, su_cold-atom_2024}:
\begin{align}\nonumber
       m_{\text{eff}} = \delta -\frac{U}{2}+2J^2 \bigg[&-\frac{4}{U + \delta} + \frac{3}{\delta - 2 U}- \frac{3}{(U - \delta)}\\\label{eq:Effective mass}
       &  + \frac{2}{\delta} - \frac{4}{(U-\delta - \eta)} \bigg].
\end{align}

\section{Experimental proposal}
\begin{figure}
    \centering
    \includegraphics[width=\linewidth]{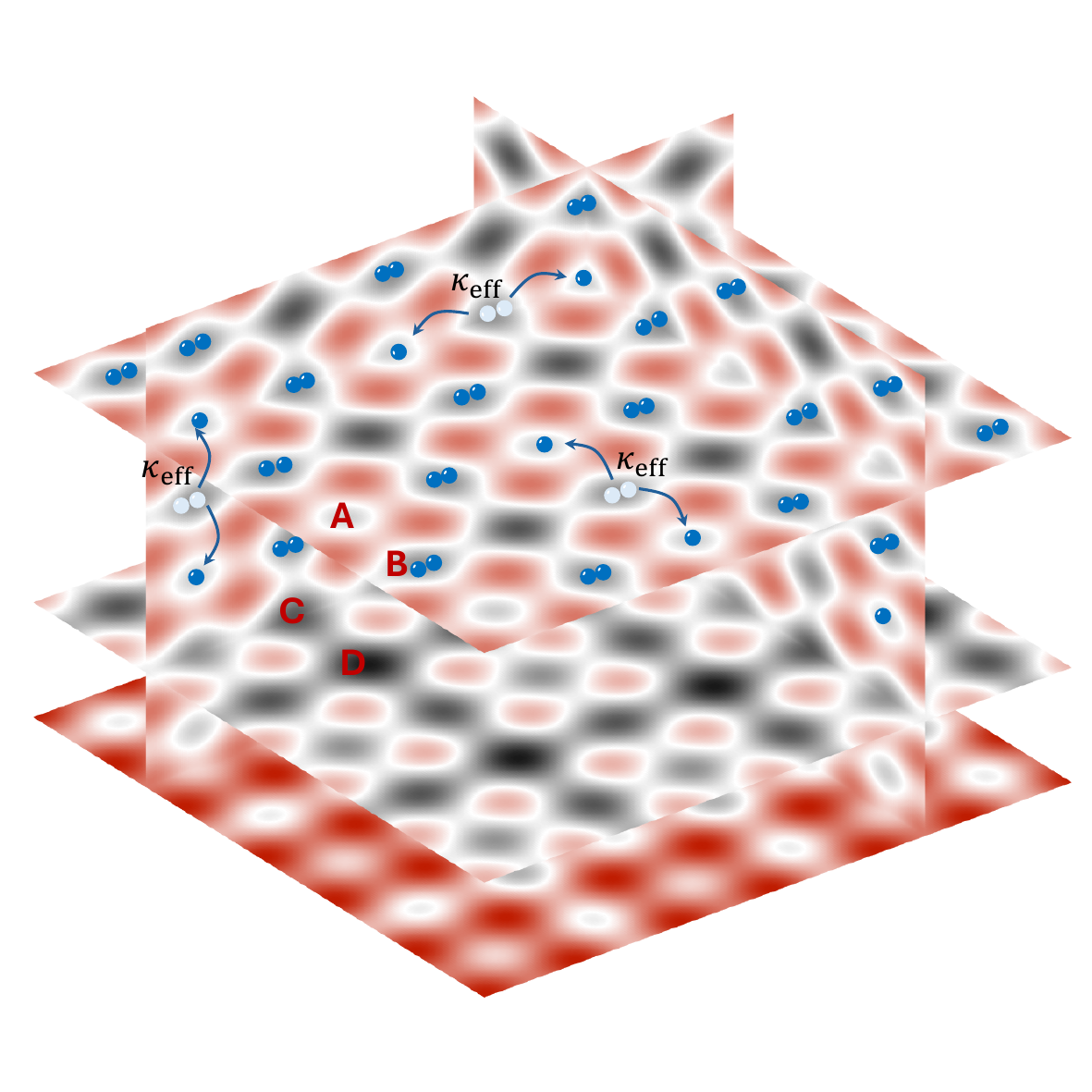}
    \caption{Illustration of a $3$d optical superlattice structure with gauge-invariant hopping along each axis. Atoms are prepared in each layer.}
    \label{fig:3dlatt}
\end{figure}

This proposal is based on an analog quantum simulation platform with ultracold atoms in the Hubbard framework~\cite{Halimeh2025}. 
Optical superlattices provide both the trapping potential and the spatially discretized lattice structure. 
The implementation can be realized using either bosonic or fermionic atoms. 
In particular, bosonic atoms offer a well-established and highly controllable experimental platform. 
A three-dimensional lattice can be constructed by applying optical superlattices along all three spatial directions. In each direction, two retro-reflected laser beams with $\lambda_l=2\lambda_s$ are superimposed to create a staggered optical super-lattice potential, with $\lambda_l$ and $\lambda_s$ being the long and short lattice potential. The potential can be written as $\sum_{i=x,y,z} V_s \cos^2{(2\pi x/\lambda_s^i)}+V_l \cos^2{(2\pi x/\lambda_l^i+\pi/4)}$. This structure naturally implements the $3+1\text{D}$ lattice gauge theory where each matter site is connected to six nearest-neighbor matter sites through six gauge links along $\pm x$, $\pm y$, and $\pm z$ directions. 

Example slices of the 3D lattice potential are illustrated in Fig.~\ref{fig:3dlatt}. The 3D superlattice structure creates local potential minima with four different depths, denoted as sites A, B, C, and D. The initial vacuum state corresponds to doublons initialized on sites B, which are gauge links. And all remaining sites are initialized empty. 

Within the Bose–Hubbard framework, a Mott insulator with uniform site occupation serves as a well-defined initial state.
We can prepare layers of $N=2$ Mott insulator sample on sites B with high fidelity using the established cooling techniques~\cite{Yang_cooling_2020}, and employing the site-dependent addressing with optical superlattices to remove undesired atomic population on sites A, C, and D~\cite{Yang2017}. 

During the subsequent time-dependent evolution, the superlattice can generate staggered potentials, and additional linear potentials can be introduced using gravity or magnetic field gradients.

In the detection stage, probing site-resolved occupations in a three-dimensional lattice is challenging due to the presence of multiple layers along the $z$ direction. 
This challenge can be addressed using a quantum gas microscope~\cite{Bakr2009,Sherson2010} together with an accordion lattice configuration~\cite{Su_Dipolar_2023}. 
The quantum state is first frozen by rapidly increasing the lattice depth along all directions.
The individual atomic layers are then separated by adjusting the interference angle of the corresponding lattice beams, increasing the interlayer spacing to several micrometers (e.g., $\sim$3~$\mu m$). 
Owing to the short depth of field of the microscope, typically about 1~$\mu m$, each layer can be imaged sequentially.
A high-precision nano-positioning stage along the $z$ direction is required to shift the objective and bring each layer into focus.
Finally, during fluorescence imaging, doublon detection can be achieved by first splitting atoms on doubly occupied sites into separate single atoms using an optical superlattice before the imaging process~\cite{yang_observation_2020}.

Another key aspect that poses a significant experimental challenge is the realization of site-dependent interactions. Such interactions can, in principle, be engineered using an optical Feshbach resonance~\cite{Theis2004}. 
In a superlattice configuration, the long-wavelength lattice beam exhibits intensity maxima at the deeper lattice sites. 
By introducing an additional laser to drive a two-photon transition to a photoassociated molecular state, the on-site interaction strength can be controlled via the frequency detuning and intensity of the light.
For a negative frequency detuning, the interaction strength at the deep lattice sites can be selectively reduced. 
However, achieving independent control of the interaction-induced energy shift while suppressing atom loss remains nontrivial. 
In practice, a careful balance must be found between tuning efficiency and loss rates associated with the optical Feshbach process.

\begin{figure}
    \centering
    \includegraphics[width=\linewidth]{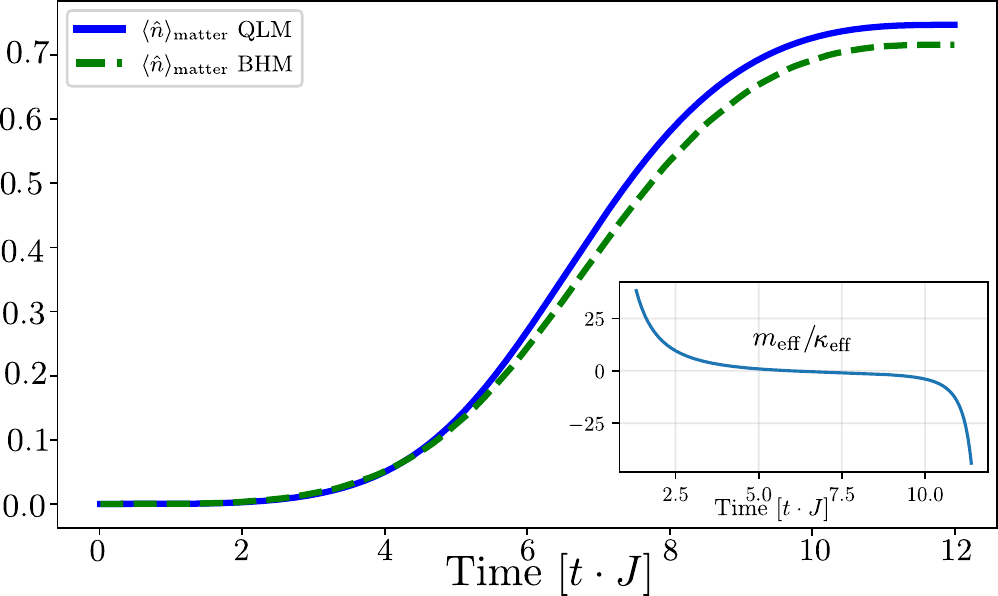}
    \caption{ Time evolution of the average matter occupation during the ramp protocol. Starting from the bare vacuum state, the system evolves into a charge-proliferated regime. The BHM hopping term $J(t)$ follows a quadratic profile, maximizing at $t=6$ where the renormalized mass vanishes ($m_{\text{ren}} = 0$), and smoothly decreasing to zero at both $t=0$ and the final time $t=t_f$. This choice of ramp profile ensures that the system begins and ends in an easily readable state, simplifying experimental protocols for measurements. The dynamics exhibits strong sensitivity to all parameters, making careful tuning essential for achieving the desired charge proliferation. The insert shows how the ratio $m_\text{eff}/\kappa_\text{eff}$ varies throughout the dynamics.}
    \label{fig:Ramp dynamics}
\end{figure}

\section{Numerical benchmarks}
For the following numerical simulations we set: $\delta = 720 $ Hz, $\boldsymbol{\gamma} = (114, 134, 940)$ Hz, $\eta = 4000$ Hz, $J=30$ Hz for the quenches, and $J_{\text{max}} = 10$ Hz for the ramp. The on-site interaction $U$ has been varied through the various simulations in order to change $m_{\text{eff}}$ as desired. 

\begin{figure}
    \centering
    \includegraphics[width=\linewidth]{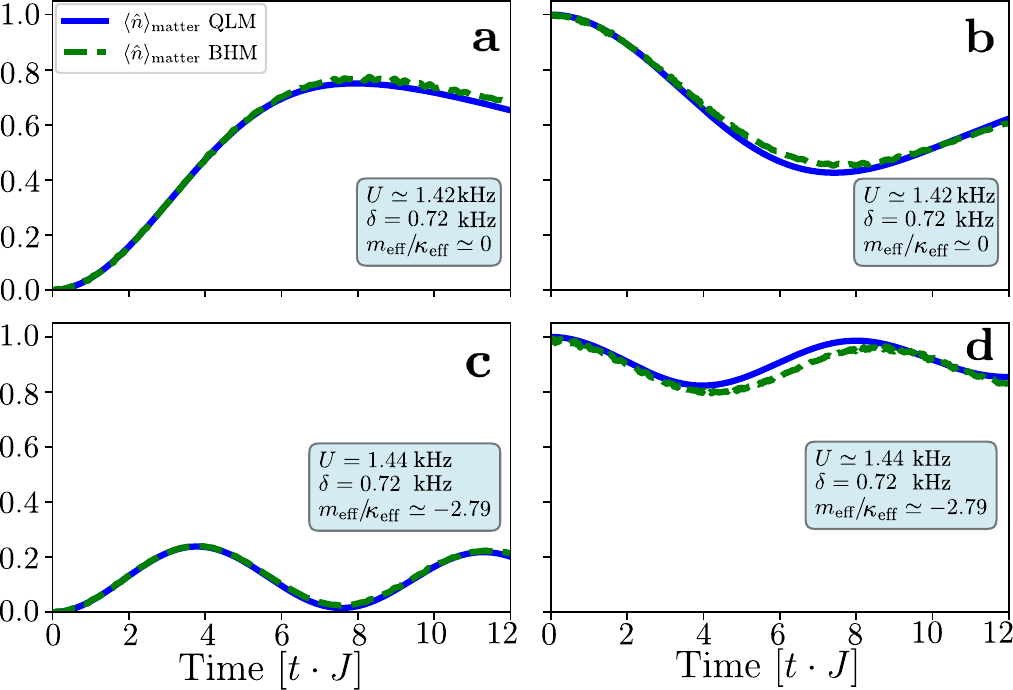}
    \caption{Time evolution of the matter occupation in the wake of global quenches, obtained using TTN simulations on an $8\times8\times4$ lattice for the BHM model, with $J = 0.03$ kHz. \textbf{a} Simulation starting from a bare vacuum state with $m_{\text{eff}} = 0$ throughout the dynamics. \textbf{b} Simulation starting from a charge proliferated state with $m_{\text{eff}} = 0$ throughout the dynamics. \textbf{c} Simulation starting from a bare vacuum state with $U = 2\delta$ which implies $m_{\text{eff}}/\kappa_{\text{eff}} \simeq -2.79$. \textbf{d} Simulation starting from a charge proliferated state with $m_{\text{eff}} = 0$ throughout the dynamics. \textbf{c} Simulation starting from charge proliferated state with $U = 2\delta$ which implies $m_{\text{eff}}/\kappa_{\text{eff}} \simeq -2.79$.}
    \label{fig:Quench_dynamics}
\end{figure}

To demonstrate the reliability of our proposed quantum simulator, we study using tree tensor networks (TTNs) \cite{shi_classical_2006, silvi_tensor_2019, felser_two-dimensional_2020, cataldi_simulating_2024, magnifico_tensor_2025, magnifico_lattice_2021} the BHM on a finite $8\times8\times4$ lattice and the corresponding target LGT. The dynamics is computed with the time-dependent variational principle (TDVP) with single tensor update\cite{haegeman_time-dependent_2011, bauernfeind_time_2020, yang_time-dependent_2020}. To run the simulation, we use \texttt{qtealeaves} library\cite{baccari_quantum_2025}.
Taking inspiration from the experiment in \cite{Yang2020ObservationGaugeInvariance}, we simulate a ramping protocol to transition from the bare vacuum state (no matter occupation) to a charge-proliferated (CP) state, see Fig.~\ref{fig:Figure1}a. The exact ramps for the relevant parameters are within experimental validity and provided in the Supplemental Material (SM) \cite{SM}. From the corresponding results in Fig.~\ref{fig:Ramp dynamics}, we can see the value of the average occupation number on matter sites over time $\langle \hat n(t)\rangle _{\rm{matter}}$ and in the inset we see how the ratio $m_{\text{eff}}/\kappa_{\text{eff}}$ between the effective mass and the effective minimal coupling varies over time. We see a transition from a vacuum state ($\langle \hat n\rangle _{\rm{matter}} = 0$) at $\m > 0$ to a charge proliferated state ($\langle \hat n\rangle _{\rm{matter}} \to  1$) at $\m <0$. We end up in a state that is not fully occupied $\langle n\rangle _{\rm{matter}}<1$ due to the finite speed of the ramp. 

To further benchmark our quantum-simulation protocol, we calculate in TTNs the dynamics under a global initial quench see Fig.~\ref{fig:Quench_dynamics}. Also in this case we can see that there is an excellent agreement between the two models, starting from either the bare vacuum or the CP state, and under different model parameters.

\begin{figure}
    \centering
    \includegraphics[width=\linewidth]{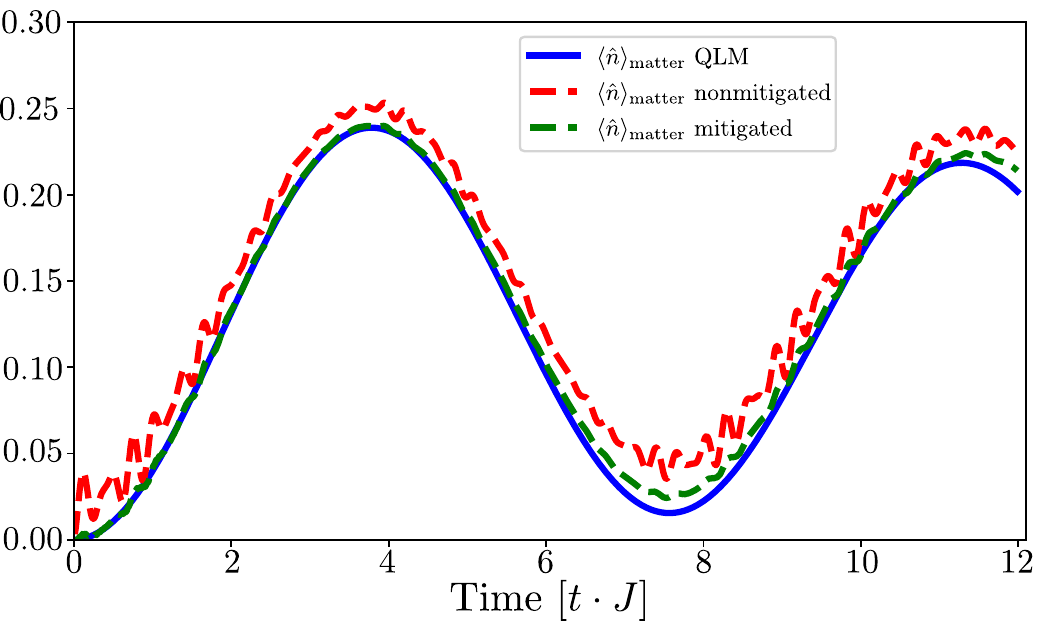}
    \caption{Comparison of the average matter occupation for raw simulation data, error-mitigated data, and the exact QLM dynamics. The system undergoes a quench from the vacuum state with an initial mass-to-hopping ratio of $m_{\text{eff}}/\kappa_{\text{eff}} = -2.79$. The BHM hopping parameter is held constant at $J = 0.03$ kHz during the time evolution. The error mitigation scheme substantially reduces deviations from the target QLM behavior, demonstrating the effectiveness of the proposed mitigation method.}
    \label{fig:Error mitigated vs normal}
\end{figure}

\section{Analog quantum error mitigation}
The results of the previous section have been further improved through a novel \textit{analog quantum error mitigation} scheme that we have developed, and which we now explain. When simulating a system by mapping it to second-order Bose--Hubbard dynamics, undesired first-order processes inevitably arise. We propose a method to remove this ``noise'' from the computed observables, thereby achieving cleaner agreement between the two models. Consider an unoccupied matter site in $1$d. To satisfy Gauss's law, it must have one empty and one doubly occupied adjacent link sites. The doubly occupied link can lead to a first-order process where one particle hops from the link to the matter site. This process is out of resonance, so at second order one of two outcomes occurs: either the particle returns from the matter site to the link (preserving the vacuum state), or the remaining particle in the link hops to the other adjacent matter site (creating an electron-positron pair). If we are in the former case, we should ignore the temporary increase in the matter field occupation. We define an observable $\hat n_s$ that counts single particles on a link. We model the error probability as the product of the probability of being in the vacuum state, $P_{\rm v} = 1-\langle \hat n\rangle_{\text{matter}}$, and the probability of having a single particle in the link, $P = \langle \hat n_s\rangle_{\text{link}}$. Subtracting this contribution from $\langle \hat n\rangle_\text{matter}$ yields an error-mitigated occupation number $\langle \hat n\rangle_{\text{EM}} = \langle \hat n\rangle_\text{matter} - (1-\langle \hat n\rangle_\text{matter})\langle \hat n_s\rangle_{\text{link}}$.
Applying the same reasoning to remove errors associated with occupied states, we obtain: $\langle \hat n\rangle_{\text{EM}} = \langle \hat n\rangle_m - (1-\langle \hat n\rangle_m)\langle \hat n_s\rangle_l + \langle \hat n\rangle_m\langle \hat n_s\rangle_l$.

In the $3$d generalization, $\langle \hat n_s\rangle_l$ acquires a factor of $3$ due to the three links per matter site. Thus:
\begin{align}\nonumber
        \langle \hat n\rangle_{\text{EM}} &= \langle \hat n\rangle_m - 3(1-\langle \hat n\rangle_m)\langle \hat n_s\rangle_l + 3\langle \hat n\rangle_m\langle \hat n_s\rangle_l \\
        &= \langle \hat n\rangle_m - 3\langle \hat n_s\rangle_l + 6 \langle \hat n\rangle_m\langle \hat n_s\rangle_l.
\end{align}
The effectiveness of this protocol is shown in Fig.~\ref{fig:Error mitigated vs normal}. For all the data shown in this work, we have used this protocol to have a better agreement with the ideal LGT.

\section{Conclusions and outlook}
We have presented a mapping of a $3+1$D U$(1)$ lattice gauge theory to a Bose--Hubbard optical superlattice in three spatial dimensions, and outlined an experimentally feasible quantum simulator with stabilized gauge symmetry through linear gauge protection induced by linear tilt potentials in $3$d. We provided a detailed experimental scheme of how this quantum simulator can be realized with existing quantum technology. We employed tree tensor network simulations to benchmark our proposed quantum simulator against the ideal target lattice gauge theory, showing excellent agreement in near- and far-from-equilibrium dynamics. We also developed an analog quantum error mitigation scheme that accounts for undesired first-order processes, further enhancing the accuracy of our quantum simulation.

Our work provides a powerful first step towards realizing quantum simulators of $3+1$D lattice gauge theories. Future avenues of exploration include going towards the lattice QED limit by employing larger values of $S$ \cite{osborne_spin-s_2023}, realizing a plaquette term \cite{Dai2017FourBodyRingExchange}, and implementing fermionic matter \cite{Surace2023AbInitioDerivation}. Aside from quantum simulation, it would also be quite interesting to explore the effect of dimensionality on the nonergodic behavior found in this model in one and two spatial dimensions.

\bigskip
\footnotesize
\begin{acknowledgments}
S.O.~and J.C.H.~acknowledge funding by the Max Planck Society, the Deutsche Forschungsgemeinschaft (DFG, German Research Foundation) under Germany’s Excellence Strategy – EXC-2111 – 390814868, and the European Research Council (ERC) under the European Union’s Horizon Europe research and innovation program (Grant Agreement No.~101165667)—ERC Starting Grant QuSiGauge. This work is part of the Quantum Computing for High-Energy Physics (QC4HEP) working group. B.Y.~acknowledges support from the NSFC (Grant No.~12274199), the Shenzhen Science and Technology Program (Grant No.~KQTD20240729102026004), the Guangdong Major Project of Basic and Applied Basic Research (Grant No.~2023B0303000011), and the Guangdong Provincial Quantum Science Strategic Initiative (Grant No.~GDZX2304006, Grant No.~GDZX2405006).
\end{acknowledgments}
\normalsize
\appendix

\section{Numerical Methods}

\begin{figure}[t]
    \centering
    \includegraphics[width=\linewidth]{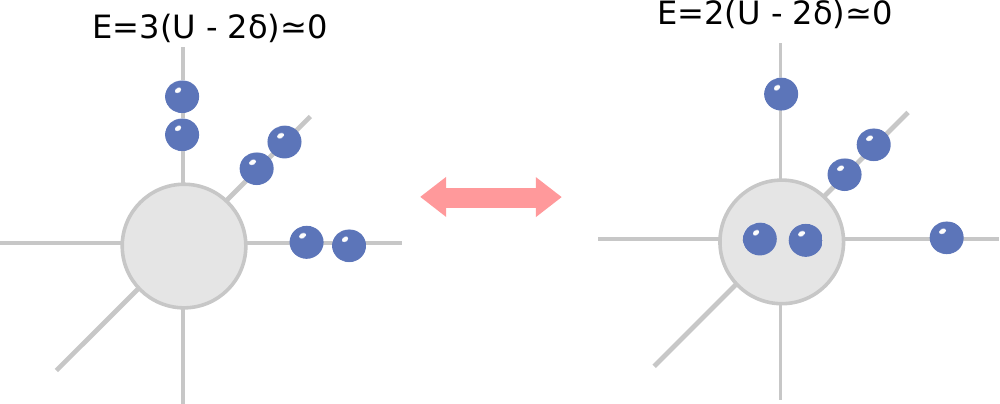}
    \caption{Pictorial representation of a process that violates the constraint of having at most one particle at a matter site in the regime $U \simeq 2\delta$ in which we work. Linear gauge protection prevents such processes.}
    \label{fig:Resonance_interaction_terms}
\end{figure}

Tensor Network (TN) methods offer an efficient framework for representing quantum many-body states by leveraging the entanglement area law characteristic of low-energy local Hamiltonians \cite{pasquale_calabrese_entanglement_2004, eisert_colloquium_2010}. 
Rather than storing the exponentially scaling coefficients of a generic wavefunction, TNs approximate the state using a polynomial number of parameters governed by the bond dimension $\chi$. 
Among TN architectures, Tree Tensor Networks (TTNs) are particularly well-suited for high-dimensional lattice models \cite{shi_classical_2006, silvi_tensor_2019, felser_two-dimensional_2020, cataldi_simulating_2024, magnifico_tensor_2025, magnifico_lattice_2021}. 
Their loopless structure enables efficient contraction with polynomial computational cost---specifically $\mathcal{O}(N d^2\chi^2 + N \chi^4)$ for binary trees\cite{qian_tree_2022}.  This favorable scaling permits the use of significantly larger bond dimensions compared to architectures like Projected Entangled Pair States (PEPS).  However, due to the lack of internal loops, TTNs do not automatically satisfy the area law in dimensions $d \geq 2$ \cite{ferris_area_2013}.  Consequently, simulation precision requires rigorous monitoring of convergence as $\chi$ increases. Beyond static optimization, TTNs effectively simulate real-time dynamics using the Time-Dependent Variational Principle (TDVP)~\cite{haegeman_time-dependent_2011, bauernfeind_time_2020, yang_time-dependent_2020}. This method projects the time-dependent Schrödinger equation onto the tangent space of the fixed-$\chi$ manifold. The approach ensures the conservation of both state norm and energy, with a computational cost per time step comparable to a standard variational ground-state sweep. The accuracy of TDVP is dictated by the time step $\delta t$ and the bond dimension $\chi$. Since entanglement grows during out-of-equilibrium evolution, $\chi$ must be sufficiently large to maintain fidelity, while $\delta t$ must be minimized to reduce projection errors.In our simulations, dynamics are initialized from a product state and evolved via single-tensor TDVP updates. To prevent variational locking and facilitate entanglement growth without determining new subspaces via SVDs, we initialize the network with a \emph{padding} scheme up to the full bond dimension $\chi$ \cite{pavesic_constrained_2025, hubig_strictly_2015, gleis_controlled_2023}. This significantly accelerates GPU performance. All reported observables have been validated for convergence with respect to both $\delta t$ and $\chi$.

\section{Site dependent interaction term}

In the regime $U\simeq 2\delta$, in $3$d, some resonances that violate our local Hilbert space restrictions arise; see Fig.~\ref{fig:Resonance_interaction_terms}. To prevent these violations, we introduced in the main text a site-dependent interaction term $U_j$ for bosons. This is related to some resonances that violate our local Hilbert space restriction in the regime $U\simeq 2\delta$ in which we work. To avoid these resonances, we use $\boldsymbol{\gamma} = (\gamma_x, \gamma_y, \gamma_z)$ where each component is different from the other two.

\bibliography{Bibiliography,biblio}

\end{document}